%% file: WheelerDC04RD1.tex
\documentclass[10pt]{article}
\usepackage{latexsym}
\usepackage{eufrak}
\usepackage{euscript}
\usepackage{graphicx}

\title{A field theoretic causal model of a Mach-Zehnder Wheeler delayed-choice experiment}

\author{ P.N. Kaloyerou\\ The University of Oxford, Wolfson College, Linton Road,\\ Oxford OX2 6UD, UK\footnote{email address: pan.kaloyerou@wolfson.ox.ac.uk}}

\newcommand{\Ab}{\mbox{\boldmath{$A$}}}
\newcommand{\Abs}{\mbox{{\scriptsize\boldmath{$A$}}}}
\newcommand{\Pb}{\mbox{{ \boldmath{$\mit \Pi$}}}}
\newcommand{\xb}{\mbox{{\scriptsize\boldmath{$x$}}}}
\newcommand{\kb}{\mbox{{\scriptsize\boldmath{$k$}}}}
\newcommand{\kbp}{\mbox{{\scriptsize\boldmath{$k'$}}}}
\newcommand{\Fi}{{\mit \Phi}}
\newcommand{\fA}{{\mit \Phi}[\Ab,t]}

\newcommand{\ep}{e^{i\kb.\xb}}
\newcommand{\epm}{e^{-i\kb.\xb}}
\newcommand{\qk}{q_{k\mu}}
\newcommand{\qks}{q_{k\mu}^{*}}
\newcommand{\pk}{\pi_{k\mu}}
\newcommand{\pks}{\pi_{k\mu}^{*}}

\newcommand{\ek}{\mbox{\boldmath $\hat\varepsilon$}_{k\mu}}
\newcommand{\ekp}{\mbox{\boldmath $\hat\varepsilon$}_{k'\mu'}}
\newcommand{\V}{V^{\frac{1}{2}}}
\newcommand{\Gd}{\mbox{\boldmath $\cal G$}}
\newcommand{\G}{\mbox{\boldmath $G$}}
\newcommand{\E}{\mbox{\boldmath $E$}}
\newcommand{\B}{\mbox{\boldmath $B$}}
\newcommand{\I}{\mbox{\boldmath $I$}}
\newcommand{\Ih}{\mbox{\boldmath $\hat I$}}
\newcommand{\epp}{e^{i(\kb-\kbp).\xb}}
\newcommand{\eppm}{e^{-i(\kb-\kbp).\xb}}
\newcommand{\ak}{\alpha_{k_\alpha\mu_\alpha}}
\newcommand{\aks}{\alpha_{k_\alpha\mu_\alpha}^{*}}
\newcommand{\bk}{\beta_{k_\beta\mu_\beta}}
\newcommand{\bks}{\beta_{k_\beta\mu_\beta}^{*}}
\newcommand{\ck}{c_{k_c\mu_c}}
\newcommand{\cks}{c_{k_c\mu_c}^{*}}
\newcommand{\dk}{d_{k_d\mu_d}}
\newcommand{\dks}{d_{k_d\mu_d}^{*}}
\newcommand{\eka}{\mbox{\boldmath $\hat\varepsilon$}_{k_\alpha\mu_\alpha}}
\newcommand{\ekb}{\mbox{\boldmath $\hat\varepsilon$}_{k_\beta\mu_\beta}}
\newcommand{\ekc}{\mbox{\boldmath $\hat\varepsilon$}_{k_c\mu_c}}
\newcommand{\ekd}{\mbox{\boldmath $\hat\varepsilon$}_{k_d\mu_d}}
\newcommand{\eko}{\mbox{\boldmath $\hat\varepsilon$}_{k_0\mu_0}}
\newcommand{\kk}{\mbox{\boldmath $k$}}
\newcommand{\ka}{\mbox{\boldmath $k$}_\alpha}
\newcommand{\kkb}{\mbox{\boldmath $k$}_\beta}
\newcommand{\kc}{\mbox{\boldmath $k$}_c}
\newcommand{\kd}{\mbox{\boldmath $k$}_d}
\newcommand{\xxb}{\mbox{\boldmath $x$}}
\newcommand{\sk}{\sum_{k\mu}}
\newcommand{\ux}{\mbox{\boldmath $u$}}
\newcommand{\vx}{\mbox{\boldmath $v$}}
\newcommand{\ffx}{\mbox{\boldmath $f$}}
\newcommand{\gx}{\mbox{\boldmath $g$}}

\begin{document}

\maketitle

\begin{abstract}
We consider a Wheeler delayed-choice experiment based on the Mach-Zehnder Interferometer. Since the development of the causal interpretation of relativistic boson fields there have not been any applications for which the equations of motion for the field have been solved explicitly. Here, we provide perhaps the first application of the causal interpretation of boson fields for which the equations of motion are solved. Specifically, we consider the electromagnetic field. Solving the equations of motion allows us to develop a relativistic causal model of  the Wheeler delayed-choice Mach-Zehnder Interferometer.  We show explicitly that a photon splits at a beam splitter. We also demonstrate the inherent nonlocal nature of a relativistic quantum field.  This is particularly revealed in a which-path measurement where a quantum is nonlocally absorbed from both arms of the interferometer. This feature explains how when a photon is split by a beam splitter it nevertheless registers on a detector in one arm of the interferometer. Bohm et al \cite{BDH85} have argued that a causal model of a Wheeler delayed-choice experiment avoids the paradox of creating or changing history, but they did not provide the details of such a model. The relativistic causal model we develop here serves as a detailed example which demonstrates this point, though our model is in terms of a field picture rather than the particle picture of the Bohm-de Broglie nonrelativistic causal interpretation.
\end{abstract}

\section{INTRODUCTION\label{SIN}}
\begin{figure}[htb]
\unitlength=1in
\hspace*{0.7in}\includegraphics[width=5in,height=3in]{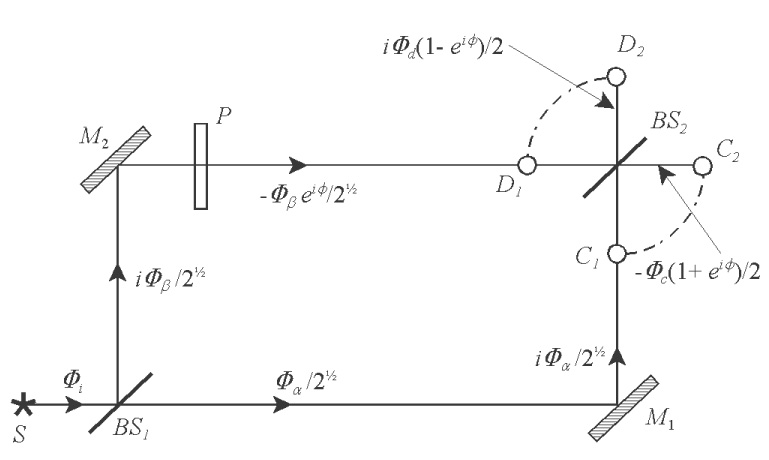}
\caption{Delayed-choice Mach-Zehnder interferometer}
\label{MZI}
\end{figure}

In  1978 Wheeler \cite{WHR78A} described seven delayed-choice experiments. The experiments are such that the choice of which complementary variable to measure is left to the last instant, long after the relevant interaction has taken place. Of the seven experiments the delayed-choice experiment based on the Mach-Zehnder interferometer is the simplest for detailed mathematical analysis. Here we present a detailed model of this experiment based on the causal interpretation of the electromagnetic field (hereafter referred to as CIEM) \cite{K94}. CIEM is a specific case of the causal interpretation of boson fields. The experimental arrangement of the delayed-choice Mach-Zehnder interferometer is shown in figure \ref{MZI}.

A single photon\footnote{Here we use the term `photon' to mean a single quantum of energy of the electromagnetic field. We do not imply any particle-like properties by this term. We have shown in reference \cite{K94A} that a single photon state is a nonlocal plane wave spread over space.} enters the interferometer at the first beam splitter $BS_{1}$. The two  beams that emerge are recombined at the second beam splitter $BS_2$ by use of the two mirrors $M_1$ and $M_2$.  $C$ and $D$ are two detectors which can be swung either behind or in front of $BS_2$. The detectors in positions $C_1$ and $D_1$ in front of $BS_2$ measure which path the photon traveled and hence a particle description is appropriate according to the orthodox interpretation. With the counters in positions $C_2$ and $D_2$  after $BS_2$ interference is observed and a wave picture is appropriate. A phase shifter $P$ producing a phase shift $\phi$ is placed in the $\beta$-beam. For $\phi=0$ and a perfectly symmetrical alignment of the beam splitters and mirrors, the $d$-beam is extinguished by interference and only the $c$-beam emerges. 
 \mbox{}\\ 

If we attribute physical reality to complementary concepts  such as wave and particle concepts, then we are forced to conclude either that (1) the history of the micro-system leading to the measurement is altered by the choice of measurement, or (2) the history of the micro-system  is created at the time of measurement.

Wheeler \cite{WHR78A} \cite{WHR83A}, following Heisenberg \cite{HEIS83}, in some sense attributed  reality to complementary concepts following measurement and adopted view (2) above, namely, that history is created at the time of measurement. Thus he states, `No phenomenon is a phenomenon until it is an observed phenomenon,' \cite{WHR78B}. He adds that `Registering equipment operating in the here and now has an undeniable part in bringing about that which appears to have happened' \cite{WHR83B}. Wheeler concludes, `There is a strange sense in which this is a ``participatory universe'' ' \cite{WHR83B}. 

In Wheeler's description, the question of the possibility of creating a causal paradox is raised. One can argue, however,  in the spirit of Bohr, that  Wheeler-delayed choice experiments are mutually exclusive in the sense that if the history of a system is fixed by one experiment, this history cannot be affected by another Wheeler delayed-choice experiment. But, it is not obvious that the paradox can be avoided in this way.  

Bohr and Wheeler share the view that `no phenomenon is a phenomenon until it is an observed phenomenon' but Bohr differs from Wheeler (and Heisenberg) in that he denies the reality of complementary concepts such as the wave concept and the particle concept. We  summerize the features of Bohr's principle of complementarity \cite{BR28}\cite{BR59A}\cite{MJA} as  follows: (1) Pairs of complementary concepts require mutually exclusive experimental configurations for their definition, (2) Classical concepts are essential as abstractions to aid thought and to communicate the results of experiment, but, physical reality cannot be attributed to such classical concepts, and (3) The experimental arrangement must be viewed as a whole, not further analyzable. Indeed, Bohr defines ``phenomenon'' to include the experimental arrangement. Hence, according to Bohr a description of underlying physical reality is impossible. It follows from this that the complementary histories leading to a measurement have no more reality than the complementary concepts to which the histories are associated. According to Bohr, then, complementary histories, like complementary concepts, are abstractions to aid thought.

In fact, Bohr had anticipated delayed-choice experiments and writes  \cite{BR59B}, `...it obviously can make no difference as regards observable effects obtainable by a definite experimental arrangement, whether our plans of constructing or handling the instruments are fixed beforehand or whether we prefer to postpone the completion of our planing until a later moment when the particle is already on its way from one instrument to another'.  Bohr also considers a Mach-Zehnder arrangement \cite{BR59C}, but not in the delayed-choice configuration. 

Complementarity is not tied to the mathematical formalism. Jammer writes  \cite{MJB}, `That complementarity and Heisenberg-indeterminacy are certainly not synonymous follows from the simple fact that the latter... is an immediate mathematical consequence of the {\it formalism} of quantum mechanics or, more precisely, of the Dirac-Jordan transformation theory, whereas complementarity is an extraneous {\it interpretative} addition to it'. Indeed, the whole process from the photon entering $BS_1$ to the final act of measurement is described uniquely by the wavefunction (or the wave functional if quantum field theory is used, as we shall see). The mathematical description leading up to the measurement is completely independent of the last instant choice of what to measure. The wave function or wave functional develops causally. Indeed, it is because in the Causal Interpretation  mathematical elements  associated with the wave function or wave functional are interpreted directly that a causal description is possible. Wheeler's assertion that a present measurement can affect the past is seen not to be a consequence of the quantum formalism, but rather, rests on an extraneous  interpretative addition. The Bohr view can also be criticized. The denial of the possibility of a description of underlying physical reality seems a high price to pay to achieve consistency.

Clearly, in a causal model of the delayed-choice experiment the issue of changing or creating history is avoided. The history leading to measurement is unique and completely independent of the last instant choice of what to measure. There is no question of a present measurement affecting the past. Bohm et al \cite{BDH85} provided just such a causal description of the Mach-Zehnder Wheeler delayed-choice experiment based on the Bohm-de Broglie causal interpretation \cite{DBR} \cite{B52}, though in general terms without solving the equations of motion. In this nonrelativistic model, electrons, protons etc. are viewed as particles guided by two real fields that codetermine each other. These are the $R$ and $S$-fields determined by the wave function $\psi(\xxb,t)=R(\xxb,t)\exp[iS(\xxb,t)/\hbar]$. The particle travels along one path which is revealed by a which-path measurement (detectors in front of $BS_2$). The $R$ and $S$-fields explain interference when the detectors are positioned after $BS_2$.

Attempts to extend the Bohm-de Broglie causal interpretation to include relativity led  to the causal interpretation of Boson fields \cite{K85}\cite{BHK87}\cite{BH93}\cite{K94} of which CIEM is a particular example.  In CIEM the beable is a field; there are no particles. Here, we apply CIEM to the Wheeler delayed-choice Mach-Zehnder interferometer. In particular, we set up and solve the equations of motion for the field. The solutions allow us to build a detailed causal model of the experiment. In CIEM the basic ontology is that of a field, not of a particle as in the Bohm-de Broglie nonrelativistic causal interpretation. We will see that a quantum field behaves much like a classical field in many respects but not in all. The essential difference is that a quantum field is inherently nonlocal. We will show explicitly that a photon is split by a beam splitter. In this case, we will have to show how in a which-path measurement, despite being split by the beam splitter, a photon registers in only one of the detectors. We will do this by modeling the detectors as hydrogen atoms undergoing the photoelectric effect, and show, using standard perturbation theory, that a photon is absorbed nonlocally from both beams by only one of the atoms. Since we have a wave model interference is explained in the obvious way. In the next section we will briefly summerize CIEM, and in section \ref{sthree} we apply CIEM to the Wheeler delayed-choice Mach-Zehnder interferometer.

\section{OUTLINE OF CIEM \label{SOOC}}
In what follows we use the radiation gauge in which the divergence of the vector potential is zero $\nabla.\Ab(\xxb,t)=0$, and the scalar potential is zero $\phi(\xxb,t) = 0$. In this gauge the electromagnetic field has only two transverse components. Heavyside-lorentz units are used throughout.

Second quantization is effected by treating the field $\Ab(\xxb,t)$ and its conjugate momentum $\Pb(\xxb,t)$ as operators satisfying  the equal-time commutation relations. This procedure is equivalent to introducing a field Schr\"{o}dinger equation
\begin{equation}
\int {\cal H} ( \Ab', \Pb') \fA\; d\xxb'= i \hbar \frac{\partial \fA}{\partial t},\label{SE}
\end{equation}
where the Hamiltonian density operator ${\cal H}$ is obtained from the classical Hamiltonian density of the  electromagnetic field,
\begin{equation}
{\cal H} =\frac{1}{2}(\mbox{\boldmath{$E$}}^{2}+\mbox{\boldmath{$B$}}^{2})=
\frac{1}{2}[c^{2}\Pb^{2}+(\nabla\times\Ab)^2 ], \label{H}
\end{equation}
by the  operator replacement $\Pb\rightarrow -i\hbar\, \delta/\delta \Ab$. $\Ab'$ is shorthand for $\Ab(\xxb',t)$ and $\delta$ denotes the variational derivative\footnote{For a scalar function $\phi$ the variational or functional derivative is defined  as $\frac{\delta}{\delta \phi}=\frac{\partial}{\partial\phi}-\Sigma_i\left(\frac{\partial}{\partial\left(\frac{\partial \phi}{\partial x_i}\right)}\right)$ \cite{SHFF68}. For a vector function $\Ab$ we have defined it to be $\frac{\delta}{\delta \Abs}=\frac{\delta}{\delta A_{x}}{\bf i}+\frac{\delta}{\delta A_{y}}{\bf j}+\frac{\delta}{\delta A_{z}}{\bf k}$.}. The solution of the field Schr\"{o}dinger  equation is the wave functional $\fA$. The square of the modulus of the wave functional $|\fA|^2$ gives the probability density  for a given field configuration $\Ab(\xxb,t)$. This suggests that we take $\Ab(\xxb,t)$ as a beable. Thus, as  we have already said, the basic ontology is that of a field; there are no photon particles.

We substitute $\Fi=R[\Ab,t]\exp(iS[\Ab,t]/\hbar)$, where $R[\Ab,t]$ and $S[\Ab,t]$ are two real functionals which codetermine one another, into the field Schr\"{o}dinger equation. Then,  differentiating, rearranging and equating imaginary  terms gives a continuity equation:
\[
\frac{\partial R^{2}}{\partial t} + c^{2} \int \frac{\delta}{\delta \Ab'}
\left(R^{2}\frac{\delta S}{\delta \Ab'} \right) \; d\xxb' = 0.
\]
The  continuity  equation is interpreted as expressing conservation of probability in function space. Equating real terms gives a Hamilton-Jacobi type equation:
\begin{equation}
\frac{\partial S}{\partial t}+\frac{1}{2}\int\left(\frac{\delta S}{\delta
\Ab'}\right)^{2} c^{2}+(\nabla\times\Ab')^{2}+\left(-\frac{\hbar^2
c^{2}}{R}\frac{\delta^{2} R}{\delta \Ab'^{2}} \right) d\xxb'= 0. \label{HJ1}
\end{equation}
This Hamilton-Jacobi equation  differs from its classical counterpart by the extra classical term 
\[
Q =-\frac{1}{2}\int\frac{\hbar^{2} c^{2}}{R}
\frac{\delta^{2} R}{\delta \Ab'^{2}}\;d\xxb', \]   
which we call the field quantum potential.
       
By analogy with classical Hamilton-Jacobi theory we define  the total energy and  momentum
conjugate to the field  as
\[
E = -\frac{\partial S[\Ab]}{\partial t},\;\;\;\;\;\Pb=\frac{\delta S[\Ab]}{\delta \Ab}.
\]

In addition to the beables $\Ab(\xxb,t)$ and $\Pb(\xxb,t)$ we can define  other field beables: the electric field, the magnetic induction, the energy and energy density, the momentum and momentum density, the intensity, etc. Formulae for these beables are obtained by replacing $\Pb$ by $\delta S/\delta \Ab$ in the classical formula. 

Thus,  we can picture an electromagnetic field as a field in the classical sense, but with the additional
property of  nonlocality. That the field is inherently nonlocal, meaning that an interaction at
one point in the field instantaneously influences the field at all other points, can be seen in
two ways: First, by using Euler's method of finite differences a functional can be
approximated as a  function of infinitely many variables:
$\fA\rightarrow\Fi(\Ab_1,\Ab_2,\ldots,t)$.  Comparison with a many-body wavefunction
$\psi(\xxb_1,\xxb_2,...,t)$ reveals the nonlocality.   The second way  is from the equation of motion
of $\Ab(\xxb,t)$, i.e.,  the free field wave equation. This is obtained by taking the  functional
derivative of the Hamilton-Jacobi equation, (\ref{HJ1}):
\[
\nabla^{2}\Ab-\frac{1}{c^{2}}\frac{\partial^{2}\Ab}{\partial t^{2}}= \frac{\delta
Q}{\delta\Ab}.
\]
In general $\delta Q/\delta\Ab$ will involve an integral over space in which the
integrand contains $\Ab(\xxb,t)$. This means that the way that $\Ab(\xxb,t)$ changes with time at
one point depends on $\Ab(\xxb,t)$ at all other points, hence the inherent nonlocality.
\subsection{Normal mode coordinates\label{NMC}}
To proceed it is mathematically easier to expand $\Ab(\xxb,t)$ and $\Pb(\xxb,t)$ as a Fourier series
\begin{equation}
\Ab(\xxb,t)=\frac{1}{\V}\sk\ek\qk(t)\ep,\;\;\;\;\;\;\;\;\;\;
\Pb(\xxb,t)=\frac{1}{\V}\sk\ek\pk(t) \epm, \label{AFS}
\end{equation}
where the field is assumed to be enclosed in a large volume $V=L^3$. The wavenumber $k$ runs from $-\infty$ to $+\infty$ and $\mu=1,2$ is the polarization index. For $\Ab(\xxb,t)$ to be a real function we must have 
\begin{equation}
\mbox{\boldmath $\hat\varepsilon$}_{-k\mu}q_{-k\mu}=\ek\qks.\label{QMP}
\end{equation}
Substituting eq.'s (\ref{H}) and (\ref{AFS})  into eq. (\ref{SE})  gives the Schr\"{o}dingier equation in terms of the normal modes $\qk$:
\begin{equation}
\frac{1}{2}\sk\left(-\hbar^{2}c^{2}\frac{\partial^2\Fi}{\partial
\qks\partial\qk }+\kappa^{2}\qks\qk\Fi\right)=
i\hbar\frac{\partial\Fi}{\partial t}. \label{SEN}
\end{equation}
The solution   $\Fi( \qk,t)$ is an ordinary
function of all the normal mode coordinates and this simplifies proceedings. 
We substitute  $\Fi=R(\qk,t)\exp[iS(\qk,t)/\hbar]$, where $R(\qk,t)$  and $S(\qk,t)$ are real functions which codetermine one another, into eq. (\ref{SEN}). Then,  differentiating, rearranging and equating real terms  gives the continuity  equation in terms of normal modes:
\[
\frac{\partial R^2}{\partial t}+\sk\left[\frac{c^2}{2}\frac{\partial}{\partial
\qk}\left(R^2\frac{\partial S}{\partial\qks}\right)+
\frac{c^2}{2}\frac{\partial}{\partial\qks}\left(R^2\frac{\partial S}{\partial\qk}
\right) \right]=0.
\]
 Equating imaginary terms gives the Hamilton-Jacobi equation in terms of normal modes:
\begin{equation}
\frac{\partial S}{\partial t}+\sk\left[\frac{c^2}{2}\frac{\partial
S}{\partial\qks}\frac{\partial S}{\partial \qk}+\frac{\kappa^{2}}{2}\qks\qk
+\left(-\frac{\hbar^{2} c^{2}}{2R}\frac{\partial^{2}R}
{\partial\qks\partial\qk}\right)\right]=0. \label{HJ2}
\end{equation}
The term 
\begin{equation}
Q = -\sk\frac{\hbar^{2} c^{2}}{2R}\frac{\partial^{2}R} {\partial\qks\partial\qk} \label{QP}
\end{equation}
is the field quantum potential. Again, by analogy with classical Hamilton-Jacobi theory we define the total energy and the conjugate momenta as
\[
E=-\frac{\partial S}{\partial t},\;\;\;\;\;\pk=\frac{\partial S}{\partial\qk},\;\;\;\;\;
\pks=\frac{\partial
S}{\partial\qks}.
\]
The square of the modulus of the wave function $|\Fi( \qk,t)|^2$  is  the probability density
for each $\qk(t)$ to take a particular value at time $t$.  Substituting a particular set of values of $\qk(t)$ at time $t$ into eq. (\ref{AFS}) gives a particular field  configuration at time $t$, as before. Substituting  the initial values of $\qk(t)$ gives the initial field configuration. 

The normalized ground state solution of the Schr\"{o}dinger equation is given by
\[
\Fi_0=N e^{-\sum_{k\mu}(\kappa/2\hbar c)q_{k\mu}^{*}q_{k\mu}}e^{-\sum_{k}i\kappa ct/2},
\]
with $N= \prod_{k=1}^{\infty}(k/\hbar c \pi)^{\frac{1}{2}}$ \footnote{The normalization factor $N$ is found by substituting $\qks=f_{k\mu}+ig_{k\mu}$ and its conjugate into $\Fi_0$ and using the normalization condition $\int_{-\infty}^{\infty} |\Fi_0|^2df_{k\mu}dg_{k\mu}=1$, with $df_{k\mu}\equiv df_{k_11}df_{k_12}df_{k_21}\ldots$, and similarly for $dg_{k\mu}$.}. Higher excited states are obtained by the action of the creation operator $a^{\dag}_{k\mu}$:
\[
\Fi_{n_{k\mu}}=\frac{(a_{k\mu}^{\dag})^{n_{k\mu}}}{\sqrt{n_{k\mu}!}}\Fi_{0}e^{- in_{k\mu}\kappa ct}.
\]
For a normalized ground state, the higher excited states remain normalized. For ease of writing we will not include the normalization factor $N$ in most expressions, but normalization of states will be assumed when calculating expectation values.

Again, the formula for the field beables are obtained by replacing the conjugate momenta $\pk$ and $\pks$ by $\partial S/\partial\qk$ and $\partial S/\partial\qks$ in the corresponding classical formula. The following is a list  of formulae for the beables:\\ \mbox{}\\
The vector potential $\Ab(\xxb,t)$ is given in eq. (\ref{AFS}). The electric field is
\begin{equation}
\E(\xxb,t)=-c\Pb(\xxb,t)=-\frac{1}{c}\frac{\partial \Ab}{\partial t}= - \frac{c}{\V}\sk\ek\frac{\partial S}{\partial\qk}\epm. \label{PEX}
\end{equation}
The magnetic induction is
\begin{equation}
\B(\xxb,t)=\nabla\times\Ab(\xxb,t)=\frac{i}{\V}\sk(\kk\times\ek)\qk(t)\ep. \label{BEX}
\end{equation}
We may also define the energy density, which includes the quantum potential density (see reference \cite{K94B}), but we will not write these here as we will not need them.  The total energy is found by integrating the energy density over $V$ to get,
\[
E=-\frac{\partial S}{\partial t}=\sk\left[\frac{c^{2}}{2} \frac{\partial
S}{\partial\qks}\frac{\partial S}{\partial\qk}+\frac{\kappa^{2}}{2}\qks\qk
+\left(-\frac{\hbar^{2}c^{2}}{2R}\frac{\partial^{2}R} {\partial
\qks\partial\qk} \right)\right]. 
\]
The intensity is equal to momentum density multiplied by $c^2$
\begin{equation}
\I(\xxb,t)=c^2\mbox{\boldmath${\cal G}$}= \frac{-ic^2}{V}\sk\sum_{k'\mu'}\left[ \ekp\times(\kk\times\ek)\frac{\partial S}{\partial q_{k'\mu'}}\qk\epp \right]. \label{I}
\end{equation}
We have adopted the classical definition of intensity in which the intensity is equal to the Poynting vector (in heavyside-lorentz units), i.e., $\I=c(\E\times\B)$. The definition leads to  a moderately simple formula for the intensity beable.  We note that the definition above contains a zero point intensity. But, because $\I$ is a vector (whereas energy is not) the contributions to the zero point intensity from individual waves with wave vector $\kk$ cancel each other because of symmetry; for each $\kk$ there is another $\kk$ pointing in the opposite direction. The above, however,  is not the definition normally used in quantum optics. This is probably because although it leads to a simple formula for the intensity beable it leads to a very cumbersome expression for the intensity operator in terms of the creation and annihilation operators:
\begin{eqnarray}
&&\mbox{{ \boldmath{$\Ih$}}}
=\frac{-\hbar c^2}{4V}\sk\sum_{k'\mu'}\left[ \frac{k}{k'}\ek\times(\kk'\times\ekp)- \frac{k'}{k}(\kk\times\ek)\times\ekp \right] \nonumber\\
&&\times\left[ \hat{a}_{k\mu}\hat{a}_{k'\mu'}e^{i(\kb+\kbp).\xb} -  \hat{a}_{k\mu}\hat{a}^\dag_{k'\mu'}\epp- \hat{a}^\dag_{k\mu}\hat{a}_{k'\mu'}\eppm +\hat{a}^\dag_{k\mu}\hat{a}^\dag_{k'\mu'}e^{-i(\kb+\kbp).\xb} \right].  \label{Ipan}
\end{eqnarray}
In quantum optics the intensity operator is defined instead as $\Ih
=c(\mbox{{ \boldmath{$\hat{E}^+$}}}\times \mbox{{ \boldmath{$\hat{B}^-$}}} - \mbox{{ \boldmath{$\hat{B}^-$}}}\times \mbox{{ \boldmath{$\hat{E}^+$}}})$, and leads to a much simpler expression in terms of creation and annihilation operators
\begin{equation}
\Ih= \frac{\hbar c^2}{V}\sk\sum_{k'\mu'}\hat{\kk}\sqrt{kk'} \hat{a}^\dag_{k\mu}\hat{a}_{k'\mu'}e^{i(\kbp-\kb).\xb}. \label{Iqo}
\end{equation}
This definition is justified because it is proportional to the dominant term in the interaction Hamiltonian for the photoelectric effect upon which instruments to measure intensity are based. We note that the two forms of the intensity operator lead to identical expectation values and perhaps further justifies the simpler definition of the intensity operator. 

From the above we see that objects such as $\qk$, $\pk$, etc. regarded as time independent operators in the Schr\"{o}dinger picture of the usual interpretation become functions of time in CIEM. 

For a given state $\Fi(\qk,t)$ of the field we determine the beables by first finding $\partial S/\partial\qk$ and its complex conjugate using the formula 
\begin{equation}
S=\left(\frac{\hbar}{2i}\right)\ln\left(\frac{\Fi}{\Fi^*}\right).\label{FlaS}
\end{equation}
This gives the beables as functions  of the $\qk(t)$ and $\qks(t)$. The beables can then be obtained in terms of the initial values by solving the equations of motion for $\qks(t)$. There are two alternative but equivalent forms of the equations of motion. The first follows from the classical formula 
\[
\pk =  \frac{\partial{\cal L}}{\partial\left(\frac{d \qk}{d t} \right)} =\frac{1}{c^2}\frac{d\qks}{d t}, 
\]
where ${\cal L}$ is the Lagrangian density of the electromagnetic field,  by replacing $\pk$ by $\partial S/\partial\qk$. This gives the equation of motion
\begin{equation}
\frac{1}{c^2}\frac{d\qks(t)}{d t} =\frac{\partial S}{\partial\qk(t)}.\label{EQMG}
\end{equation}
 
The second form of the equations of motion for $\qk$ is obtained by differentiating the Hamilton Jacobi equation (\ref{HJ2}) by $\qks$. This gives the wave equations
\begin{equation}
\frac{1}{c^2}\frac{d^{2}\qks}{d t^{2}}+\kappa^{2}\qks=-\frac{\partial
Q}{\partial\qk}. \label{WEQ}
\end{equation}
The corresponding equations   for $\qk$ are the complex conjugates of the above. These equations of motion  differ from the classical free field wave equation by the derivative of the quantum potential. From this it follows that where the quantum potential is zero or small the quantum field  behaves like a classical field. In applications we will obviously choose to solve the simpler eq. ({\ref{EQMG}}). 

In the next section we apply CIEM to the Mach-Zehnder Wheeler delayed-choice experiment.

\section{A CAUSAL MODEL OF THE MACH-ZEHNDER WHEELER DELAYED-CHOICE EXPERIMENT\label{sthree}}
Consider the Mach-Zehnder arrangement shown in figure \ref{MZI}. $BS_1$ and $BS_2$  are beam splitters, $M_1$ and $M_2$ are mirrors and $P$ is a phase shifter that shifts the phase of a wave by an amount $\phi$. In what  follows we will assume for simplicity that the beam suffers a $\pi/2$ phase shift at each reflection and a zero phase shift upon transmission through a beam splitter. In general, phase shifts upon reflection and transmission may be more complicated than this. The requirement is that the commutation relations must be preserved. The latter is a more stringent requirement than energy conservation (or, equivalently, of photon number conservation) alone \cite{CST}. The polarization unit vector  is unchanged by either reflection or transmission.

\subsection{Input Region}
The input region is the region before the first beam splitter $BS_1$. The incoming beam is a Fock state containing one quantum:
\[
\Fi_i(\qk,t) =\left(\frac{2\kappa_0}{\hbar c}\right)^{\frac{1}{2}}
q_{k_0\mu_0}^{*}(t)\Fi_{0}e^{-i\kappa_0 ct}.
\]
From the formula  (\ref{FlaS}) we can find the $S$ corresponding to the state $\Fi_i(\qk,t)$. Using this result in eq. (\ref{EQMG}) gives the equations of motion:
\begin{equation}
\frac{dq_{k_0\mu_0}^{*}}{dt}=c^2\frac{\partial S}{\partial q_{k_0,\mu_0}}  =\frac{i\hbar c^2}{2q_{k_0\mu_0}}, \;\;\;\;\;\;\;\;\;\;\frac{d\qks}{dt} =c^2 \frac{\partial S}{\partial \qk}=0\;\;\; \mathrm{for}\;\;\; k\neq \pm k_0. \label{DS0}
\end{equation}
The solutions are easily found:
\begin{equation}
q_{k_0\mu_0}^{*}(t)=q_0e^{i(\omega_0 t +\theta_0 )},\;\;\;\;\;\;\;\;\;\;\qks(t)= q_{k\mu 0}e^{i\zeta_{k\mu 0}}\;\;\; \mathrm{for}\;\;\;k\neq \pm k_0. \label{Sol0}
\end{equation}
$q_0$ and $q_{k\mu 0}$ are constant amplitudes, and $\theta_0$  and $\zeta_{k\mu 0}$ are constant phases all fixed at $t=0$. These initial conditions cannot of course be precisely determined but are given with some probability found from $|\Fi_i(...q_{k\mu}^{*}..., ...q_{k\mu},t)|^2$. 

We find expressions for the beables in the input region by substituting the $\partial S/\partial q_{k_0,\mu_0}$ and $\partial S/\partial \qk$ found from eq. (\ref{DS0}) together with  solutions (\ref{Sol0}) into  the formulae for the beables. In finding the expression for the beables we use formula ({\ref{QMP}). Defining $\Theta_0=k_0.\xxb-\omega_0 t -\theta_0$ the beables are:
\begin{eqnarray}
\Ab(\xxb,t)&=&\frac{2}{\V}\eko q_0\cos\Theta_0+\frac{\gx_A(\xxb)}{\V}, \nonumber\\
\E(\xxb,t)&=&\frac{-2\omega_0}{\V c}q_0\eko \sin\Theta_0\label{E0},\\
\B(\xxb,t)&=&\frac{-2}{\V}q_0(k_0\times \eko)\sin\Theta_0+ \frac{\gx_B(\xxb)}{\V},\nonumber
\end{eqnarray}
with 
\[
\gx_A(\xxb)= \sum_{\stackrel{\scriptstyle{k\mu}}{k\neq \pm k_0}}\ek\qk\ep, \;\;\;\;\;\;\;\;\;\; \gx_B(\xxb)=\nabla\times\gx_A(\xxb)= i\sum_{\stackrel{\scriptstyle{k\mu}}{k\neq \pm k_0}}(k\times\ek)\qk\ep. 
\]

\subsection{Region I \label{R1}}
We consider the state $\Fi_{I}$ in region I and determine from this state the corresponding beables. Region  I is the region after the phase shifter $P$ and the mirror $M_1$ but before $BS_2$. 

The incoming beam  $\Fi_{i}$  is split at $BS_{1}$ into two beams: the $\alpha$ and 
$\beta$-beams\footnote{Some workers insist that two inputs into the beam splitter must be used even when one of the inputs  is the vacuum \cite{OHM87}, while other workers use a single input \cite{L73} \cite{SZ97}. In passing, we mention  that Caves \cite{C80}, in connection with the search for gravitational waves  using a Michelson interferometer, suggests, as one of two possible explanations, that vacuum fluctuations are responsible for the `standard quantum limit' which places a limit on the accuracy of any measurement of the position of a free mass. We will use the single input approach for $BS_1$ as it simplifies the mathematical analysis while all essential results remain the same.}.  The $\alpha$-beam undergoes a $\pi/2$ phase shift at $M_{1}$ and becomes $\Fi_{\alpha}e^{i\pi/2}=i\Fi_{\alpha}$. The $\beta$-beam undergoes two $\pi/2$ phase shifts followed by a $\phi$  phase shift and becomes $\Fi_{\beta}e^{i\phi}e^{i\pi} =-\Fi_{\beta}e^{i\phi}$. Also multiplying by a $1/\sqrt{2}$ normalization factor the state $\Fi_{I}$ in region I becomes
\begin{equation}
\Fi_{I}=\frac{1}{\sqrt{2}}\left( i\Fi_{\alpha}-\Fi_{\beta}e^{i\phi}\right), \label{PHRI}
\end{equation}
where $\Fi_{\alpha}$ and $\Fi_{\beta}$ are solutions of the normal mode Schr\"{o}dinger equation (\ref{SEN}) and are given by
\[
\Fi_{\alpha}(\qk,t) = \left(\frac{2\kappa_\alpha}{\hbar c}\right)^{\frac{1}{2}}
\aks\Fi_{0}e^{-i\kappa_\alpha ct}, \;\;\;\;\;\;\;\;\;
\Fi_{\beta}(\qk,t) = \left(\frac{2\kappa_\beta}{\hbar c}\right)^{\frac{1}{2}}
\bks\Fi_{0}e^{-i\kappa_\beta ct}.
\]
The magnitudes of the $k$-vectors are equal, i.e., $k_\alpha=k_\beta=k_0$. 

By using formula (\ref{FlaS}) to first determine $S$, 
\begin{equation}
S=\frac{\hbar}{2i}\left[ -\sum_k ikct-2ik_0 ct +\ln\left( i\aks-\bks e^{i\phi}\right)- \ln\left(- i\ak-\bk e^{-i\phi}\right) \right], \label{SS} 
\end{equation}
we can determine $\partial S/\partial\ak$, $\partial S/\partial\bk$, and $\partial S/\partial\qk$. Substituting the latter into eq. (\ref{EQMG}) gives the equations of motion for $\aks$, $\bks$ and $\qks$:
\begin{eqnarray}
\frac{d\aks}{dt}&=&c^2\frac{\partial S}{\partial\ak}=\frac{-\hbar c^2}{2}\frac{1}{\left[i\ak+\bk e^{-i\phi}\right]},   \label{EQMa} \\
\frac{d\bks}{dt} &=&c^2\frac{\partial S}{\partial\bk} =\frac{-\hbar c^2}{2i}\frac{e^{-i\phi}}{\left[i\ak+\bk e^{-i\phi}\right]}, \label{EQMb}\\
\frac{d\qks}{dt} &=&c^2\frac{\partial S}{\partial\qk}  =0,\;\;\;\; \mathrm{for}\; k\neq \pm k_{\alpha}, \pm k_{\beta}.  \label{EQMq}
\end{eqnarray}
Eq.'s (\ref{EQMa}) and (\ref{EQMb}) are coupled differential equations. The coupling shows that the $\alpha$ and $\beta$-beams depend nonlocally on each other. Taking the ratio of Eq.'s (\ref{EQMa}) and (\ref{EQMb}) gives the relation
\begin{equation}
\aks(t)=ie^{i\phi}\bks(t), \label{RAB}
\end{equation}
which can be used to decouple the two differential equations. The decoupled differential equations and eq. (\ref{EQMq}) are easily solved to give
\begin{equation}
\aks(t)=\alpha_0 e^{i(\omega_\alpha t+\sigma_0)}, \;\;\;
\bks(t)=\beta_0 e^{i(\omega_\beta t+\tau_0)}, \;\;\;
\qks(t)= q_{k\mu 0}e^{i\zeta_{k\mu 0}}\;\mathrm{for}\;k\neq \pm k_{\alpha}, \pm k_{\beta},\label{QKS}
\end{equation} 
where $\sigma_0$ and $\tau_0$ are integration constants corresponding to the initial phases, and $\alpha_0$ and $\beta_0$ are constant amplitudes. Different values of the constants  $q_{k\mu 0}$ and 
$\zeta_{k\mu 0}$ in different regions of the interferometer would correspond to amplitude and phase changes of the normal mode coordinates of the ground state. These changes to the normal mode coordinates of the ground state, if indeed they occur, do not lead to any observable differences in measured physical quantities so that there is no way (at present) to detect them. For this reason we choose $q_{k\mu 0}$ and $\zeta_{k\mu 0}$ to have the same values in all the regions of the interferometer. The omega's, $\omega_\alpha=\hbar  c^2/4\alpha_0^2$ and $\omega_{\beta}=\hbar c^2/4\beta_0^2$, are nonclassical frequencies which depend on the amplitudes $\alpha_0$ and $\beta_0$.

Substitution eq.'s (\ref{QKS}) into eq. (\ref{RAB}) gives the following relations among the constants associated with $\ak(t)$ and $\bk(t)$:
\begin{equation}
\alpha_ 0=\beta_0,\;\;\;\;\;\;\;\;\;\;\;\;\;\;\;\;\;\;\;\sigma_0=\tau_0+\phi+\frac{\pi}{2}.\label{IABR}
\end{equation}
Substituting eq.'s (\ref{IABR}) into $\omega_\alpha$ or $\omega_\beta$ shows that $\omega_\alpha=\omega_\beta$. 
\subsection{The beables in region I \label{BRI}}
In this section we obtain explicite expressions for the beables $\Ab(\xxb,t)$, $\E(\xxb,t)$, $\B(\xxb,t)$ and $\I(\xxb,t)$ as functions of time and the initial values. The expression for the energy density is very cumbersome and is not as useful in the present context as the intensity. For this reason we will  not give the energy density here. For the same  reason we will also leave out the quantum potential density, though we will need the quantum potential. We note that in what follows $\eko=\eka=\ekb$, where $\eko$ is the polarization of the incoming wave. 
 
To find the beables as explicite functions of position and time we substitute the partial derivatives of $S$ with respect to the normal mode coordinates found in eq.'s (\ref{EQMa}), (\ref{EQMb}), and (\ref{EQMq})  together with the solutions for $\ak(t)$, $\bk(t)$ and $\qk(t)$ given in eq.'s (\ref{QKS}) into the formulae for the beables given  in eq.'s (\ref{AFS}), (\ref{PEX}), (\ref{BEX}), and (\ref{I}). Eq. (\ref{QMP}) is used in deriving the beable expressions. After lengthy manipulation  and simplification, and defining $\Theta_{\alpha}=\ka.\xxb-\omega_{\alpha} t-\sigma_0$ and $\Theta_{\beta}=\kkb.\xxb-\omega_{\beta} t-\tau_0$ we get:
\begin{eqnarray}
\Ab_I(x,t)&=&\frac{2}{\V}\left(\eka\alpha_0\cos\Theta_{\alpha}+\ekb\beta_0\cos\Theta_{\beta}
\right) +\frac{\ux_I(\xxb)}{\V},\nonumber \\ 
\E_I(\xxb,t)&=&\frac{-\hbar c}{2\V}\left(\frac{\eka}{\alpha_0}\sin\Theta_{\alpha}+\frac{\ekb}{\beta_0}\sin\Theta_{\beta} \right), \label{EI}\\
\B_I(\xxb,t) &=&\frac{-2}{\V}\left[(\ka\times\eka)\alpha_0\sin\Theta_{\alpha} + (\kkb\times\ekb)\beta_0 \sin \Theta_{\beta}\right]  + \frac{\vx_I(\xxb)}{\V}, \nonumber\\ 
\I_I(\xxb,t)&=&\frac{\hbar c^2}{2V}\left(\ka+\kkb-\ka\cos2\Theta_{\alpha} -\kkb\cos 2\Theta_{\beta}\right)-\frac{\ffx_I(\xxb)\gx_I(\xxb,t)}{V}, \nonumber
\end{eqnarray}
with
\begin{eqnarray}
\ux_I(\xxb)&=&\!\!\!\!\!\sum_{\stackrel{\scriptstyle{k\mu}}{k\neq \pm k_{\alpha},\pm k_{\beta}}}\!\!\!\!\!\ek\qk\ep, 
\;\;\;\;\;\vx_I(\xxb)=\nabla\times\ux_I(\xxb)=i\!\!\!\!\!\sum_{\stackrel{\scriptstyle{k\mu}}{k\neq \pm k_{\alpha},\pm k_{\beta}}}\!\!\!\!\!(\kk\times\ek)\qk\ep, \label{VXI}\\
\ffx_I(\xxb)&=&i\hbar c^2\!\!\!\!\!\sum_{\stackrel{\scriptstyle{k\mu}}{k\neq \pm k_{\alpha},\pm k_{\beta}}}\!\!\!\!\!\eka\times(\kk\times\ek)\qk\ep,\;\;\;\;\;
\gx_I(\xxb,t)=\sin \Theta_{\alpha}+\sin \Theta_{\beta}.\nonumber
\end{eqnarray}

What the above beables show, and the point we want to emphasize, is that the single input photon is split by the beam splitter $BS_1$ into two beams. Each beam carries half the total momentum,
\[
\G_I= \int_V \Gd_I dV=\int_V \frac{\I_I}{c^2}\;dV= \frac{\hbar \ka}{2}+\frac{\hbar \kkb}{2}=\hbar  k_0,
\]
and energy. The energy distribution given by the Hamilton-Jacobi equation (\ref{HJ2}) turns out to be very cumbersome to calculate, but the total energy is easily found from eq. (\ref{SS}):
 \[
E =-\frac{\partial S}{\partial t}=\hbar c k_0+ \sum_k \frac{\hbar c k}{2}.
\]
There is no question of the whole photon choosing a single  path through the beam splitter. We note that since $\Fi_I$ is an eigenstate of the energy and momentum operators, the expectation values of these operators are equal to the energy and momentum beables.

We see that the field beable behaves much like a classical electromagnetic field, but there are two differences: The first difference is that frequencies $\omega_{\alpha}=\omega_{\beta}$ have a nonclassical dependence on the amplitude of the field. The second and more significant difference is that each beam depends nonlocally on the other beam. This is shown, as we have already mentioned, by the fact that the equations of motion of the two beams, eq. (\ref{EQMa}) and eq. (\ref{EQMb}), are  coupled differential equations. That is to say, the behaviour of each beam depends nonlocally on the behaviour of the other beam.

The nonlocal dependence of each beam on the other can also be seen from the wave equations  of $\ak$ and $\bk$. These can be found by inserting the total quantum potential in region I (found by using formula (\ref{QP})),
\[
Q_I=-\frac{1}{2}\sk k^2\qks\qk+\hbar c k_0+\sum_k \frac{\hbar c k}{2} -\frac{\hbar^2 c^2}{2h_I^* h_I},
\]
into the wave equation (\ref{WEQ}) and differentiating. This gives
\begin{eqnarray}
\frac{1}{c^2}\frac{d^2\ak}{dt^2} &=& \frac{-\hbar^2 c^2 \left[ \ak-i\bk e^{-i\phi}\right]}{2(h_I^*h_I)^2},   \nonumber\\
\frac{1}{c^2} \frac{d^2\bk}{dt^2}&=& \frac{-\hbar^2 c^2\left[\bk+i\ak e^{i\phi} \right]}{2(h_I^* h_I)^2}, \nonumber
\end{eqnarray}
with $h_I=-i\ak-\bk\exp(-i\phi)$. In each wave equation the right hand side depends on functions from both beams and therefore indicates a nonlocal time dependence of each beam on the other.

Using eq. (\ref{VXI}) it is easy to show that the above expressions for the $\Ab_I(\xxb,t)$, $\E_I(\xxb,t)$, and $\B_I(\xxb,t)$ beables satisfy the usual classical relations $\E=-(1/c)\partial \Ab_I/\partial t$ and $\B_I=\nabla\times\Ab_I$.

There are a number of ways to establish a connection between the initial (constant) amplitudes and phases of the input region and region I. One convenient way is to compare the electric field beables in the $\alpha$ and $\beta$-beams, eq. (\ref{EI}), with  those obtained by beginning with the input electric field beable, eq. (\ref{E0}), and inserting the appropriate amplitude and phase changes as it splits at $BS_1$ and then passes through  $M_1$, $M_2$ and the phase shifter $P$ into region I. This comparison gives the relation between the initial constants in the input region and in region I:
\[
\alpha_0=\beta_0=\frac{q_0}{\sqrt{2}},\;\;\;\;\;\;\;\;\;\;\;\; 
\sigma_0=\theta_0-\frac{\pi}{2},\;\;\;\;\;\;\;\;\;\;\;\; 
\tau_0=\theta_0-\phi-\pi. 
\]

\input{wheelerdc04RD2}

\end{document}

%% file: wheelerdc04RD2.tex
\subsection{Which-path measurement}
From the above we see that the beam is split into two parts. With the detectors positioned before $BS_2$ we know that only one detector will register the absorption of a quantum of electromagnetic energy. We must now show how this comes about even though the incoming quantum of electromagnetic energy is split into two parts.

To do this we consider an idealized measurement using the photoelectric effect for a position measurement. We model the detectors in positions  $C_1$ and $D_1$ in figure \ref{MZI} as hydrogen atoms in their ground state. We assume that the incoming electromagnetic quantum has sufficient energy to ionize a hydrogen atom. Each beam interacts with one of the hydrogen atoms. To see what will happen we will focus on the interaction of the $\alpha$-beam with the hydrogen atom at position $C_1$. 

We will treat the hydrogen atom nonrelativistically and picture it as made up of a proton and an electron particle according to the Bohm-de Broglie nonrelativistic ontology.  From the perspective of the description of the electromagnetic field, this nonrelativistic approximation of the atom compared to a relativistic treatment will involve only a minor differences in accuracy, but will not alter in anyway the model of the field in the interaction process. In the authors opinion, a satisfactory relativistic fermion ontology has not yet been achieved. A fully relativistic treatment may therefore involve a change in the ontology of fermions that we have assumed here. 

The interaction of the electromagnetic field with a hydrogen atom is described by the Schr\"{o}dinger equation
\[
i\hbar\frac{\partial \Fi}{\partial t}= (H_R+H_A+H_I)\Fi. 
\]
$H_R$, $H_A$ and $H_I$ are the free radiation, free atomic, and interaction Hamiltonians, respectively, and are given by
\[
H_R = \sk\left(a^{\dag}_{k\mu}a_{k\mu}+\frac{1}{2}\right)\hbar\omega_k, \;\;\;\;\; H_A = \frac{-\hbar^2}{2\mu}\nabla^2+V(\xxb), \;\;\;\;\;H_I = \frac{i\hbar e}{\mu c}\left(\frac{\hbar c}{2V}\right)^{\frac{1}{2}} \sk\frac{1}{\sqrt{k}}a_{k\mu}\ep\ek.\nabla,
\]
with $\omega_k=kc$ and $\mu=m_em_n/(m_e+m_n)$ is the reduced mass. The interaction Hamiltonian is derived, as usual,  using the Pauli minimal coupling. Only the first order terms are retained. We also drop the term in $H_I$ containing the creation operator since  a hydrogen atom in its ground state cannot emit a photon.

 The initial combined state of the radiation and atom is
\[
\Fi_{I_{k\mu}i}(\qk,\xxb,t)=\Fi_{I_{k\mu}}(\qk)u_i(\xxb)e^{-ik_\alpha ct}e^{-\sum_k ikct/2}e^{-iE_{ei}t/\hbar}.
\]
We take the initial state of the radiation $\Fi_{I_{k\mu}}$ to be eq. (\ref{PHRI}) with the subscript $I$ replaced by $I_{k\mu}$. Th initial state of the hydrogen atom is taken to be its ground state,
\[
u_i(\xxb,t)=\frac{1}{\sqrt{\pi a^3}}e^{r/a}e^{-iE_{ei}t/\hbar},
 \]
where $a=4\pi\hbar^2/\mu e^2$ is the Bohr magneton. 

We assume a solution of the form 
\begin{equation}
\Fi(\qk,\xxb,t)=\sum_{N_{k\mu}}\sum_{n}[a^{(0)}_{N_{k\mu}n}(t)+ a^{(1)}_{N_{k\mu}n}(t)] \Fi_{N_{k\mu}}(\qk)u_n(\xxb) e^{-iE_Nt/\hbar}, \label{ASL}
\end{equation}
where we have retained only the zeroth and first order expansion coefficients, and where $E_N=E_{N_{k\mu}}+E_{en}$.  
$E_{N_{k\mu}}$ is the energy of the field including the zero point energy $E_0$.  $E_{en}=\hbar k_0 c-E_{ei}$ is the kinetic energy of the liberated electron and  $E_{ei}$ is the ionization energy of the atom. $\Fi_{N_{k\mu}}$ is the final number (Fock) state, and $u_n(\xxb)=\exp(i \kb_{en}.\xb)/\sqrt{V}$ is the outgoing normalized electron plane wave.To find the expansion coefficients we use the formulae below derived using standard perturbation theory:
\begin{eqnarray}
a^{(0)}_{N_{k\mu}n}(t)&=&\delta_{N_{k\mu}I_{k\mu}}\delta_{ni},\label{ao}\\
a^{(1)}_{N_{k\mu}n}(t) &=& H_{N_{k\mu}n, I_{k\mu}i} \frac{1}{i\hbar}\int^{t}_{0}
e^{i\omega_{N_{k\mu}n,I_{k\mu}i}t}\;dt, \label{FOEC}\\
H_{N_{k\mu}n,I_{k\mu}i}&=& \int\Fi^*_{N_{k\mu}n}(\qk,\xxb)H_I \Fi_{I_{k\mu}i}(\qk,\xxb)d\qk d\xxb\label{ME},\\
\hbar\omega_{N_{k\mu}n,I_{k\mu}i}&=&E_{N_{k\mu}n}-E_{I_{k\mu}i} = E_{N_{k\mu}}+E_{en}-E_{I_{k\mu}}-E_{ei}=E_{N_{k\mu}n,I_{k\mu}i}.\label{ENA}
\end{eqnarray}
Evaluating the time integral in eq. (\ref{FOEC}) gives 
\begin{equation}
\frac{1}{i\hbar}\int^{t}_{0}e^{i\omega_{N_{k\mu}n,I_{k\mu}i}t}\;dt =\frac{1-e^{iE_{N_{k\mu}n,I_{k\mu}i}t/\hbar}}{E_{N_{k\mu}n,I_{k\mu}i}}. \label{TI}
\end{equation}
The modulus squared of eq. (\ref{TI}) in the limit  $t\rightarrow\infty$ becomes $
\delta(E_{N_{k\mu}n,I_{k\mu}i})2\pi t/\hbar$. This corresponds to energy conservation. We require the time $t$ over which the integral is taken to be very much longer than $\hbar/E_n$, but  sufficiently short that $a^{(0)}(t)$ does not change very much. In this case, it is a good approximation to take $t\rightarrow\infty$ as we have done above. 

After substituting for $H_I$ in eq. (\ref{ME}) the matrix element becomes
\begin{equation}
H_{N_{k\mu}n,I_{k\mu}i}=\frac{-e}{\mu c}\sqrt{\frac{\hbar c}{2V}} \left[\sum_{k\mu}\frac{1}{\sqrt{k}}\int \Fi^{*}_{N_{k\mu}}(\qk)a_{k\mu}\Fi_{I_{k\mu}}(\qk) \;d\qk\right]
.\left[\int u^{*}_n(\xxb) \ep\ek.(-i\hbar\nabla) u_i(\xxb)\;d\xxb\right]. \label{HNI}
\end{equation}
The first integral is
\begin{equation}
\sum_{k\mu}\frac{1}{\sqrt{k}}\int\Fi^{*}_{N_{k\mu}}a_{k\mu}\Fi_{I_{k\mu}}\;d\qk 
= \frac{1}{\sqrt{2k_0}}(i-e^{i\phi})\int \Fi^{*}_{N_{k\mu}}\Fi_0\;d\qk= \frac{1}{\sqrt{2k_0}}(i-e^{i\phi})\delta_{N_{k\mu}0}\delta_{kk_0}\delta{\mu\mu_0}. \label{FIAKMU}
\end{equation}
Using the dipole approximation $\ep=1$ and the fact that the delta functions $\delta_{kk_0}\delta{\mu\mu_0}$ in the first integral produce $\ek\rightarrow\eko$, the second integral of eq. (\ref{HNI}) is evaluated to give
\begin{equation}
\int u^{*}_n(\xxb)\eko.(-i\hbar\nabla) u_i(\xxb)\;d\xxb=\frac{\hbar}{\sqrt{V\pi a^3}}\eko.\kk_{en} \frac{8\pi a^3}{(1+a^2 k_{en}^2)^2}, \label{IUSU}
\end{equation}
where $\eko.\kk_{en}$ is the component of the outgoing electron wave vector in the direction of polarization $\eko$ of the incident electromagnetic field. 

We draw attention to the fact that the  integral (\ref{FIAKMU}) demonstrates that if interaction takes place at all then the atom must absorb one entire quantum and leave the field in its ground state. Substituting eq.'s (\ref{ao}), (\ref{FOEC}), (\ref{TI}), (\ref{HNI}),  (\ref{FIAKMU}) and (\ref{IUSU}) into the assumed solution eq. (\ref{ASL}) gives the final solution
\begin{equation}
\Fi=\Fi_{I_{k\mu}i}(\qk,\xxb,t)-\frac{\Fi_0(\qk,t)}{V}\sum_n \eta_{0n}(t)\eko.\kk_{en}\frac{1}{\sqrt{V}}e^{i\left(\kb_{en}.\xb-E_{en}t/\hbar\right)}, \label{FSOL}
\end{equation}
with
\[
\eta_{0n}(t)=\left(\frac{e}{\mu c}\right)\sqrt{\frac{\hbar c}{2V}}\left[\frac{(i-e^{i\phi})}{\sqrt{2 k_0}}\right]\left[\frac{\hbar}{\sqrt{V\pi a^3}} \frac{8\pi a^3}{(1+a^2 k_{en}^2)^2} \right]\left(\frac{1- e^{iE_{0n,I_{k\mu}i}t/\hbar}}{E_{0n,I_{k\mu}i}}\right).
\]
$E_{0n,I_{k\mu}i}$ is given by eq. (\ref{ENA}) with $N_{k\mu}=0$.

In the solution (\ref{FSOL})  the first term is the initial state and corresponds to no interaction taking place. Recall that for a single atom the probability of interaction with the electromagnetic field is small. The second term shows that if the interaction takes place then {\it one entire} quantum must be absorbed. This means that the {\it field energy from both beams is absorbed by only one of the hydrogen atoms}. Since the arms of the interferometer can be of arbitrary length, and since the duration of the interaction with the atom is small, the absorption of the electromagnetic quantum occurs nonlocally. This is the second way in which nonlocality enters into the description, and further emphasizes the difference from a classical field. In this way, we show how despite the fact that a single quantum is necessarily split by the first beam splitter, only one of the counters placed before $BS_2$ fires.

\subsection{Region II}
With the counters in the $C_2$ and $D_2$ positions interference is observed. Since we are modeling the quantum mechanical electromagnetic field as a real field, interference is explained in the obvious way even though the field has additional nonclassical properties. Here we want to show the interference of the beables explicitely by solving the equations of motion for the normal mode coordinates. 

At $BS_2$ the $\alpha$ and $\beta$-beams are split forming the $c$ and $d$-beams by interference (see figure \ref{MZI}). We call the region after $BS_2$ region II. We want to find the state $\Fi_{II}$ of the field in this region and to determine the corresponding beables.

The part of the $\alpha$-beam  reflected at $BS_2$ suffers another  $\pi/2$ phase shift. The part of the $\beta$-beam transmitted at $BS_2$ suffers no phase shift. Each beam is again multiplied by $1/\sqrt{2}$ because  of the 50\% intensity reduction at $BS_2$. The two beams interfere to form the $c$-beam represented by the state $-(1/2)\Fi_{c}(1+e^{i\phi})$. 

The transmitted part of the $\alpha$-beam experiences no phase change, while the part of the $\beta$ beam that  is reflected at $BS_2$ suffers a further $\pi/2$ phase change. These two beams interfere to form the $d$-beam represented by the state $(i/2)\Fi_d(1-e^{i\phi})$. Adding these two states gives the state of the field in region $II$
\[
\Fi_{II}=-\frac{1}{2}\Fi_{c}(1+e^{i\phi})+\frac{i}{2}\Fi_d(1-e^{i\phi}).
\]
$\Fi_c$ and $\Fi_d$ are Fock states given by
\[
\Fi_c(\qk,t) = \left(\frac{2\kappa_c}{\hbar c}\right)^{\frac{1}{2}}
\cks(t)\Fi_{0}e^{-i\kappa_c ct}, \;\;\;\;\;\;\;\;\;\;\Fi_d(\qk,t) = \left(\frac{2\kappa_d}{\hbar c}\right)^{\frac{1}{2}} \dks(t)\Fi_{0}e^{-i\kappa_d ct}. 
\]
Note that the magnitudes of the $k$-vectors are unchanged by interaction with optical elements, i.e., $k_c=k_d=k_\alpha=k_\beta=k_0$. As before, by using formula (\ref{FlaS}) to determine $S$,
\begin{eqnarray}
S&=&\frac{\hbar}{2i}\left\{-\sum_k ikct -2ik_0 ct +\ln\left[-(1+e^{i\phi})\cks +i(1-e^{i\phi})\dks\right]\right. \nonumber\\
&& \left.  -\ln\left[-(1+e^{-i\phi})\ck-i(1-e^{-i\phi})\dk\right]\right\},\nonumber
\end{eqnarray}
we can determine the derivatives of $S$ with respect to the normal mode coordinates. Substituting the latter into eq. (\ref{EQMG}) gives the equations of motion of the normal modes:
\begin{eqnarray}
\frac{d\cks}{dt}&=&c^2 \frac{\partial S}{\partial\ck}=\frac{c^2\hbar }{2} \frac{(1+e^{-i\phi})}{\left[-i(1+e^{-i\phi})\ck+(1-e^{-i\phi})\dk \right]},  \nonumber\\ 
\frac{d\dks}{dt} &=&c^2 \frac{\partial S}{\partial\dk} =\frac{c^2\hbar}{2} \frac{i(1-e^{-i\phi})}{\left[-i(1+e^{-i\phi})\ck+(1-e^{-i\phi})\dk\right]}, \nonumber \\ 
\frac{d\qks}{dt} &=& c^2 \frac{\partial S}{\partial\qk}  =0,\;\;\;\; \mathrm{for}\; k\neq \pm k_c, \pm k_d  \label{EQMQII}
\end{eqnarray}
By taking the ratio of the two coupled differential equations to get the  relation 
 \begin{equation}
i(1-e^{-i\phi})\cks=(1+e^{-i\phi})\dks,\label{RCD}
\end{equation}
and by following similar steps to those of section (\ref{R1}), we can find the  solutions of the equations of motion (\ref{EQMQII}):
\begin{equation}
\cks(t)=c_0 e^{i(\omega_c t+\chi_0)},\;\;\;
\dks(t)=d_0 e^{i(\omega_d t+\xi_0)}, \;\;\;
\qks(t)= q_{k\mu 0}e^{i\zeta_{k\mu 0}} \;  \mathrm{for}\; k\neq \pm k_c, \pm k_d, \label{CDST}
\end{equation}
with $\omega_c=[\hbar c^2(1+\cos\phi)]/4c_0^2$ and $\omega_d=[\hbar c^2(1-\cos\phi]/4d_0^2$. The $c_0$, $d_0$ and $q_{k\mu 0}$ are constant amplitudes, and $\chi_0$, $\xi_0$ and the $\zeta_{k\mu 0}$ are constant phases. 

Setting $t=0$ in eq.'s (\ref{RCD}), and (\ref{CDST}) and solving the resulting equations gives the relation between the initial (constant) values in region II:
\begin{equation}
d_ 0=-c_0 \tan\frac{\phi}{2}, \;\;\;\;\;\;\;\;\;\;\;\;\;\;\;\;\;\;\; \chi_0=\xi_0.  \label{ICDR}
\end{equation}
Substituting eq.'s (\ref{ICDR}) into $\omega_c$ or $\omega_d$ shows that $\omega_c=\omega_d$. 

\subsection{The beables in region II} 
As in section \ref{BRI}, to get expressions for the beables we substitute the derivatives of $S$ with respect to the normal mode coordinates given in eq.'s (\ref{EQMQII}) together with the solutions for the normal mode coordinates, eq.'s (\ref{CDST}),  into the formulae for the beables given in subsection \ref{NMC}.  We note that the polarization remains unchanged in interactions with optical elements, i.e., $\ekc=\ekd=\eka=\ekb=\eko$ and that we have used eq. (\ref{QMP}). After lengthy manipulation and simplification, and defining $\Theta_c=\kc.\xxb-\omega_c t-\chi_0$ and $\Theta_d=\kd.\xxb-\omega_d t -\xi_0$ we get:
\begin{eqnarray}
\Ab_{II}(x,t)&=&\frac{2}{\V}\left( \ekc c_0 \cos\Theta_c +\ekd d_0\cos \Theta_d\right)+\frac{\ux_{II}(\xxb)}{\V},\nonumber\\
\E_{II}(\xxb,t)&=&\frac{-\hbar c}{2\V}\left(\frac{\eka}{c_0}(1+\cos\phi)\sin\Theta_c + \frac{\ekd}{d_0}(1-\cos\phi)\sin\Theta_d \right),\nonumber\\
\B_{II}(\xxb,t) &=&\frac{-2}{\V}\left[(\kc\times\ekc)c_0\sin\Theta_c + (\kd\times\ekd)d_0 \sin\Theta_d\right] +\frac{\vx_{II}(\xxb)}{\V}, \nonumber\\
\I_{II}(\xxb,t)&=&\frac{\hbar c^2}{2V}\left[\kc(1+\cos\phi)+\kd(1-\cos\phi) -\kc(1+\cos\phi)\cos2\Theta_c+ \kd(1-\cos\phi)\cos 2\Theta_d)\right] \nonumber\\
&&-\frac{\ffx_{II}(\xxb)\gx_{II}(\xxb,t)}{V}, \label{INTII}
\end{eqnarray}
with
\begin{eqnarray}
\ux_{II}(\xxb)&=&\!\!\!\!\!\sum_{\stackrel{\scriptstyle{k\mu}}{k\neq \pm k_c,\pm k_d}}\!\!\!\!\!\ek\qk\ep,\;\;\;\;\;
\vx_{II}(\xxb)=i\!\!\!\!\!\sum_{\stackrel{\scriptstyle{k\mu}}{k\neq \pm k_c,\pm k_d}}\!\!\!\!\!(\kk\times\ek)\qk\ep=\nabla\times \ux_{II}(\xxb), \label{VXII}\\
\ffx_{II}(\xxb)&=&\frac{i\hbar c^2}{V}\!\!\!\!\!\sum_{\stackrel{\scriptstyle{k\mu}}{k\neq \pm k_c,\pm k_d}}\!\!\!\!\!\eko\times(\kk\times\ek)\qk\ep,\;
\gx_{II}(\xxb,t)=(1+\cos\phi)\sin\Theta_c +(1-\cos\phi)\sin\Theta_d.\nonumber
\end{eqnarray}

Using eq. (\ref{VXII}) we can again show that the above expressions for the $\Ab_{II}(\xxb,t)$, $\E_{II}(\xxb,t)$, and $\B_{II}(\xxb,t)$ beables satisfy the usual classical relations $\E_{II}=-(1/c)\partial\Ab/\partial t$ and $\B_{II}=\nabla\times\Ab_{II}$. 

By tracing the input electric field beable, eq. (\ref{E0}), through all of the optical elements and then comparing with the electric field beable obtained from the equations of motion (as we did in section \ref{BRI}) we obtain relations among the initial (constant) values in all three regions:
\begin{eqnarray}
c_0&=&\sqrt{2}\alpha_0\cos\frac{\phi}{2}, \;\;\;\;\;\;\;\;\;\;\;\; 
d_0=-\sqrt{2}\alpha_0\sin\frac{\phi}{2}, \;\;\;\;\;\;\;\;\;\;\;\; 
\chi_0=\xi_0=\sigma_0-\frac{\phi}{2}-\frac{\pi}{2}. \nonumber\\
c_0&=&q_0\cos\frac{\phi}{2}, \;\;\;\;\;\;\;\;\;\;\;\;\;\;\;\;\;\;
d_0=-q_0\sin\frac{\phi}{2}, \;\;\;\;\;\;\;\;\;\;\;\;\;\;\;\;\;\;\;\; 
\chi_0=\xi_0=\theta_0-\frac{\phi}{2}-\pi. \label{ICDR2}
\end{eqnarray}
Substituting eq. (\ref{ICDR2}) into  $\omega_c$ or $\omega_d$ shows that $\omega_0=\omega_c=\omega_d=\omega_\alpha=\omega_\beta$.

The beables above clearly exhibit interference. For example, with a phase shift of $\phi=0$ the $d$-beam is extinguished, while for $\phi=\pi$ the $c$-beam is extinguished. For $\I_{II}(\xxb,t)$ and the $\E_{II}(\xxb,t)$ beables this is obvious, but for $\Ab_{II}(x,t)$ and $\B_{II}(\xxb,t) $ we need to use relation (\ref{ICDR}).

In the experiment what is actually observed is the expectation value of the intensity operator, eq. (\ref{Ipan}) (or eq.(\ref{Iqo}) which gives the same results). This expectation value is equal to the long time average (which is equal to the cycle average for a periodic function) of the intensity beable, eq.(\ref{INTII}), i.e.,
\[
\left\langle \I_{II}\right\rangle_{cycle} = \frac{1}{T_c}\int^{T_c/2}_{-T_c/2}\I_{II} \;dt= \frac{\hbar c^2}{2V}[\kc(1+\cos \phi)+\kd(1-\cos \phi)],
\]
with $T_c=2\pi/\omega_c$. The total momentum and energy of each beam,
\[
\G_{II}=\int_V \frac{\I_{II}}{c^2}\;dV= \frac{\hbar}{2V}[\kc(1+\cos \phi)+\kd(1-\cos \phi)]=\hbar k_0,\;\;\;\;\; E =\hbar c k_0+\sum_k \frac{\hbar c k}{2},
\]
are again equal to the corresponding expectation values, as we would expect.

As with the $\alpha$ and $\beta$-beams the two coupled differential equations in (\ref{EQMQII})  show that the two beams are nonlocally connected. This is further demonstrated by the wave equations
\begin{eqnarray}
\frac{1}{c^2}\frac{d^2\ck}{dt^2} &=& \frac{-2\hbar^2 c^2\left[ (1+\cos\phi)\ck+\sin\phi\;\dk\right]}{(h_{II}^*h_{II})^2}, \nonumber \\
\frac{1}{c^2} \frac{d^2\dk}{dt^2}&=& \frac{-2\hbar^2 c^2\left[\sin\phi\;\ck+(1-\cos\phi)\dk\right]}{(h_{II}^*h_{II})^2}, \nonumber
\end{eqnarray}
obtained by substituting the quantum potential in region II,
\[
Q_{II}=-\frac{1}{2}\sk k^2\qks\qk+\hbar c k_0 +\sum_k \frac{\hbar k c}{2} -\frac{\hbar^2 c^2}{h_{II}^* h_{II}},
\]
with $h_{II}=-[1+\exp(-i\phi)]\ck-i[1-\exp(-i\phi)]\dk$,  into the wave equation (\ref{WEQ}).

\section{CONCLUSION}
We have provided perhaps one of the first applications of the causal interpretation of relativistic boson fields where the equations of motion  for the field are solved explicitly. In so doing we have been able to provide a detailed relativistic causal model of the Wheeler delayed-choice Mach-Zehnder Interferometer, i.e., a description of the physical reality that underlies the experiment. We have shown explicitly that a single photon is split by a beam splitter.  We have shown that the beables representing quantities such as the electric field and the magnetic induction behave much like their classical counterparts. One similarity is that the expressions for the electric field and magnetic induction beables in terms of the vector potential beable are the same as for their classical counterparts.  They differ from their classical counterparts  in that the beables oscillate with  a nonclassical frequency which depends  on the amplitude of the wave. A more significant difference is the inherent nonlocality of a relativistic quantum field. This nonlocality is revealed in two ways: First, because the time dependence of a beam in one arm of the interferometer depends nonlocally on the time dependence of the beam in the other arm. Second, in a which-path measurement an entire quantum is absorbed nonlocally from both arms of the interferometer by a detector placed in one arm. This feature explains how when a photon is split by a beam splitter it nevertheless registers in a detector placed in one arm of the interferometer.  Wheeler concludes from his hypothetical delayed-choice experiments that history is created at the time of measurement. Bohr tells us that complementary concepts such as wave and particle concepts are abstractions to aid thought, and hence, so also is any historical evolution  leading to the final experimental result. We have argued that neither Wheeler's nor Bohr's conclusions follow from the mathematical formalism of the quantum theory. The Bohm-de Broglie nonrelativistic causal interpretation and its relativistic generalization to boson fields demonstrates that we neither need to follow Bohr and deny  a description of underlying physical reality, nor do we need to follow Wheeler and conclude that the present can affect the past.

%% file: WheelerDC04RD1.bbl
\begin{thebibliography}{99}
\bibitem{BDH85}  D. Bohm,  C. Dewdney, and B.J. Hiley,  Nature {\bf 315}, 294 (1985).
\bibitem{WHR78A}  J.A. Wheeler, in {\it Mathematical Foundations
of Quantum Theory} edited by E.R. Marlow (Academic Press, 1978), p. 9.
\bibitem{K94} P.N. Kaloyerou, Phys. Rep. {\bf 244}, 287 (1994).
\bibitem{K94A} Ref. \cite{K94} p. 327.
\bibitem{WHR83A}J.A. Wheeler in {\it Quantum Theory and Measurement}, edited by J.A. Wheeler and W.H. Zurek (Princeton University Press, 1983), p.182.
\bibitem{HEIS83}W. Heisenberg in {\it Quantum Theory and Measurement}, edited by J.A. Wheeler and W.H. Zurek (Princeton University Press, 1983), p. 62.
\bibitem {WHR78B} Ref. \cite{WHR78A} p. 14.
\bibitem {WHR83B} Ref. \cite{WHR83A} p. 194.
\bibitem{BR28} N. Bohr at {\it Atti del Congresso Internazionale dei Fisici}, Como, 11-20 September 1927 (Zanichelli, Bologna, 1928), Vol. 2   p. 565; substance of the Como lecture is reprinted in Nature {\bf 121}, 580 (1928) and in N. Bohr, {\it Atomic Theory and the Description of Nature} (Cambridge University Press, Cambridge, 1934) p. 52. 
\bibitem{BR59A} N. Bohr in {\it Albert Einstein: Philosopher Scientist}, 
edited by P.A. Schilpp (Open Court Publishing Company, third edition, 1982) p. 201.
\bibitem{MJA} M. Jammer, {\it The Philosophy of Quantum Mechanics: The
Interpretations of Quantum Mechanics in Historical Perspective} (John Wiley \&
Sons, 1974).
\bibitem{BR59B} Ref. \cite{BR59A} p. 230.
\bibitem{BR59C} Ref. \cite{BR59A} p. 222.
\bibitem{MJB}Ref. \cite{MJA} p. 60.
\bibitem{DBR}  L. de Broglie,  {\it Une Tentative d'Interpr\'{e}tation Causale et
Non-Lin\'{e}aire de la M\'{e}canique Ondulatoire (La Th\'{e}orie de la Double Solution)} (Gauthier-Villars, Paris, 1956); english translation:   {\it Non-linear Wave Mechanics, a Causal Interpretation} (Elsevier, Amsterdam,1960).
\bibitem{B52} D. Bohm,  Phys. Rev., {\bf 85}, 166 (1952) ; {\bf 85}, 180 (1952).
\bibitem{K85} P.N. Kaloyerou, Ph.D. Thesis, University of London, 1985.
\bibitem{BHK87} D. Bohm, B.J. Hiley, and P.N. Kaloyerou, Phys. Rep. {\bf 144}, 349 (1987).
\bibitem{BH93} D. Bohm and B.J. Hiley,  {\it The Undivided Universe: an
Ontological Interpretation of Quantum Theory} (Routledge, 1993).
\bibitem{SHFF68} L.I. Schiff, {\it Quantum Mechanics} (McGraw-Hill Kogakusha, third edition, 1968), p. 494.
\bibitem{K94B} Ref. \cite{K94} p. 317.
\bibitem{CST} R.A. Campos, B.E.A. Saleh, and M.C. Teich, Phys. Rev. A {\bf 40}, 1371 (1989), p. 1373.
\bibitem{OHM87} Z.Y. Ou, C.K. Hong and L. Mandel, Opt. Commun. {\bf 63}, 118 (1987). 
\bibitem{L73} R. Loudon, {\it The Quantum Theory of Light} (Oxford University Press, second edition, 1983) p. 222. Here the beam splitter described as part of the Hanbury Brown and Twiss experiment.
\bibitem{SZ97} M.O. Scully and M. S. Zubairy, {\it Quantum Optics} (Cambridge University Press, 1997) p. 494. Here the beam splitter is used as part as an atomic interferometer.
\bibitem{C80} C.M. Caves, Phys. Rev. Lett. {\bf 45}, 75 (1980).

\end{thebibliography}
